\documentclass[reprint,superscriptaddress,floatfix,longbibliography]{revtex4-2}
\usepackage{blindtext}
\usepackage{amsfonts}
\usepackage{amsmath}
\usepackage{amssymb,bbm}
\usepackage{graphicx}%
\usepackage{color}
\usepackage{epstopdf}
\usepackage {multirow}

\usepackage[capitalize]{cleveref}
\usepackage{bm}%
\usepackage[LGR,T1]{fontenc}
\usepackage[latin9]{inputenc}
\usepackage{array}
\usepackage{booktabs}
\usepackage{textcomp}
\usepackage{multirow}
\usepackage{diagbox}
\usepackage{braket}
\usepackage{url}
\makeatletter

\ProvideTextCommand{\~}{LGR}[1]{\char126#1}

\makeatother

\DeclareSymbolFont{myletters}{OML}{ztmcm}{m}{it}
\DeclareMathSymbol{\uplambda}{\mathord}{myletters}{"15}
\begin{document}
	\title{Generalized Spin Groups and Applications to Magnons}
	\author{Xinhai Tu}
	\affiliation{National Laboratory of Solid State Microstructures and School of Physics, Nanjing University, Nanjing 210093, China}
	\affiliation{Collaborative Innovation Center of Advanced Microstructures, Nanjing University, Nanjing 210093, China}
	\author{Di Wang}
	\affiliation{National Laboratory of Solid State Microstructures and School of Physics, Nanjing University, Nanjing 210093, China}
	\affiliation{Collaborative Innovation Center of Advanced Microstructures, Nanjing University, Nanjing 210093, China}
	\author{Yuxuan Mu}
    \affiliation{National Laboratory of Solid State Microstructures and School of Physics, Nanjing University, Nanjing 210093, China}
    \affiliation{Collaborative Innovation Center of Advanced Microstructures, Nanjing University, Nanjing 210093, China}	
	\author{Xiangang Wan}
	\email{xgwan@nju.edu.cn}
	\affiliation{National Laboratory of Solid State Microstructures and School of Physics, Nanjing University, Nanjing 210093, China}
	\affiliation{Collaborative Innovation Center of Advanced Microstructures, Nanjing University, Nanjing 210093, China}
	\affiliation{Hefei National Laboratory, Hefei 230088, China}
	\affiliation{Jiangsu Physical Science Research Center, Nanjing University, Nanjing 210093, China}
	\begin{abstract}
Spin groups provide the natural symmetry framework for the isotropic Heisenberg model, in which spin and lattice degrees of freedom are separable. They have recently found broad applications in describing electronic band structure, magnon band topology, and transport phenomena. However, a typical spin Hamiltonian describing magnons contains not only isotropic Heisenberg exchange but also anisotropic interactions induced by spin-orbit coupling (SOC), including the Dzyaloshinskii-Moriya (DM) interaction, the $\Gamma$ exchange, and the single-ion anisotropy (SIA), which strongly influence magnon properties. More recently, some studies have also revealed relativistic magnon transport phenomena in altermagnets. Such systems typically exhibit partial spin-lattice locking, but are commonly assumed to be described by conventional magnetic groups. Here, we develop a systematic symmetry framework, generalized spin groups, for magnons. We first classify the symmetries of Heisenberg, DM, $\Gamma$, and SIA interactions. The resulting classification contains 6 spin-only groups, $\mathbb Z_{2}^{\mathcal S}\otimes \mathbb{SO}(3)^{\mathcal S}$, $\mathbb Z_{2}^{\mathcal S}\otimes \mathbb{SO}(2)^{\mathcal S}$, $\mathbb Z_{2}^{\mathcal S}\otimes \mathbb{O}(2)^{\mathcal S}$, $\mathbb Z_{2}^{\mathcal S}\otimes \mathbb{D}_{2}^{\mathcal S}$, $\mathbb Z_{2}^{\mathcal S}\otimes \mathbb{C}_{2}^{\mathcal S}$, and $\mathbb Z_{2}^{\mathcal S}$, together with a total of 161 nontrivial spin groups. The symmetries of magnetically ordered phases are then obtained through spontaneous symmetry breaking of the corresponding parent groups for spin Hamiltonians. We also identify 6 spin-only groups for ordered states, $\mathbb{C}_{\infty v}$, $\mathbb{SO}(2)$, $\mathbb{C}_{s}$, $\mathbb{C}_{2v}$, $\mathbb{C}_{2}$, and $\mathbbm{1}$, which generate 90, 264, 157, 264, 387, and 598 nontrivial spin groups, respectively. Finally, we apply generalized spin groups to the magnon thermal conductivity and spin conductivity tensors and derive all the symmetry-allowed forms. Our results provide a unified symmetry framework for analyzing magnon transport both with and without SOC, yielding more refined constraints on magnon responses.
	\end{abstract}
	\maketitle
\textit{Introduction.--}The symmetry analysis of magnetic materials has traditionally been formulated within the framework of magnetic groups \cite{Shubnikov1964Colored,Bradley1968Magnetic}, where time-reversal symmetry $T$ is broken and spin transformations are assumed to be locked to spatial operations by spin-orbit coupling (SOC). This framework has been remarkably successful in classifying magnetic structures \cite{Opechowski1965Magnetic,Birss1964Symmetry} and constraining response functions \cite{Kleiner1966Transport,PhysRevB.92.155138}. However, early studies of spin-wave spectra revealed degeneracies beyond those expected from magnetic-group symmetry alone in magnets well described by the isotropic Heisenberg model with negligible SOC \cite{Harris1963YIG,Dimmock1962MagneticCrystals}. To resolve this problem, Brinkman and Elliott introduced spin groups \cite{doi:10.1098/rspa.1966.0211,10.1063/1.1708514}, in which spin-space and real-space operations are treated as separable symmetry operations. Litvin and Opechowski \cite{LITVIN1974538,Litvin1977} later reformulated this idea in a rigorous group-theoretical language and classified all nontrivial spin point groups. More recently, the complete classifications of spin space groups for collinear, coplanar, and noncoplanar magnetic configurations have also been obtained \cite{PhysRevX.14.031037,PhysRevX.14.031038,PhysRevX.14.031039}. The spin-group framework has since been applied to magnon band topology \cite{Corticelli2022SpinSpaceMagnonTopology,Chen2025}, electronic band structures \cite{PhysRevX.12.021016}, and transport phenomena \cite{PhysRevX.15.031006}.

However, a typical spin Hamiltonian contains not only isotropic Heisenberg exchange but also SOC-induced anisotropic interactions, most notably the bilinear spin interactions including the antisymmetric Dzyaloshinskii-Moriya (DM) interaction \cite{DZYALOSHINSKY1958241,PhysRev.120.91}, the symmetric $\Gamma$ exchange \cite{PhysRev.120.91,PhysRevB.108.085140} and the single-ion anisotropy (SIA) \cite{White2007}. These interactions can qualitatively reshape magnon band structures and modify their properties. DM interactions can gap magnon band crossings and generate finite Berry curvature, thereby giving rise to transverse magnon transport, including the magnon thermal Hall effect \cite{Katsura2010ThermalHall,Onose2010MagnonHall,Matsumoto2011RotatingMagnon,Mook2014TopologicalMagnonInsulator,Laurell2018MagnonThermalHall} and the magnon spin Nernst effect \cite{PhysRevLett.117.217202,PhysRevLett.117.217203,PhysRevB.96.134425}. Symmetric anisotropic exchange constrained by crystal symmetry can induce nonreciprocal magnon propagation and contribute to topological magnon bands \cite{Matsumoto2020NonreciprocalMagnons,McClarty2018TopologicalMagnons}. SIA selects easy-axis or easy-plane moment orientations and controls low-energy anisotropy gaps in the magnon spectrum \cite{Liu2022SIA,Scheie2023NiPS3}. The symmetry of such systems is conventionally described by magnetic groups, which encode fully locked spin and spatial transformations. Recent studies, however, have identified special settings in which spin-group symmetries can remain exact or serve as useful approximations even in the presence of SOC \cite{PhysRevX.14.031037,1qd6-ddzl,Etxebarria:cam5007}. Nevertheless, these results do not provide a general theoretical foundation, nor do they establish complete symmetry constraints on magnon transport. This motivates a systematic symmetry theory for magnons beyond the isotropic Heisenberg limit.
\begin{figure*}
	\centering
	\includegraphics[width=15cm]{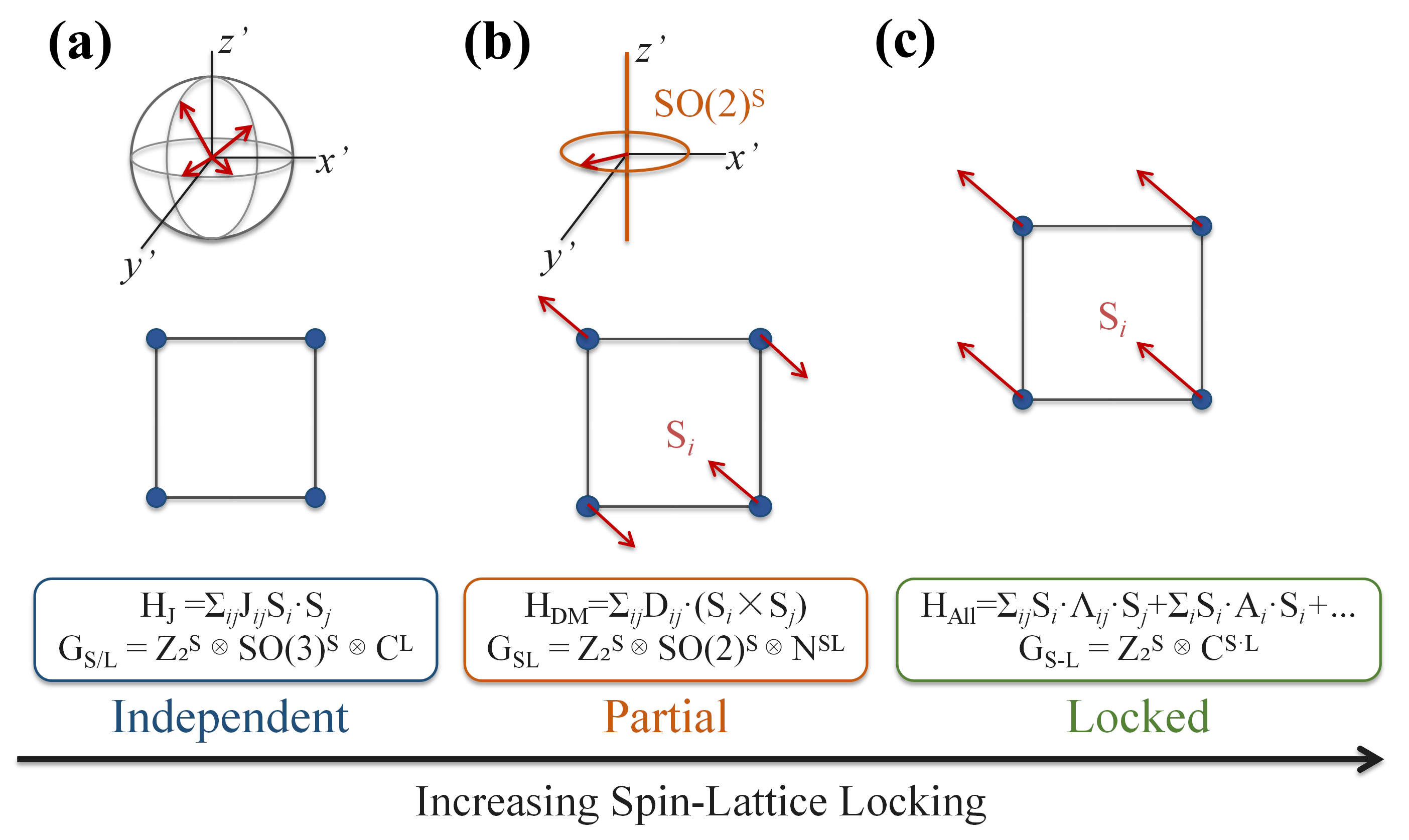}
	\caption{Schematic progression from independent to fully locked spin and lattice operations. (a) In the absence of SOC, spin operations of spin Hamiltonian act independently of lattice operations , and the symmetry of the Heisenberg Hamiltonian is described by $\mathbb Z_2^{\mathcal S}\otimes \mathbb{SO}(3)^{\mathcal S}\otimes \mathbb C^{\mathcal L}$. (b) With DM interaction, residual spin-only operations coexist with combined spin-lattice operations. This partially locked spin Hamiltonian can be described by GSGs. (c) In the limiting case of complete spin-lattice locking, spin operations are tied to lattice operations, and the symmetry of the spin Hamiltonian has symmetry $\mathbb Z_2^{\mathcal S}\otimes \mathbb C^{\mathcal S\cdot\mathcal L}$.}\label{fig1}
\end{figure*}

In this Letter, we develop generalized spin groups (GSGs) as a unified framework for the symmetry analysis of magnons. Each GSG contains a spin-only group (SOG), formed by transformations acting exclusively in spin space, together with a nontrivial spin group (NSG) that combines spin and spatial transformations. We first determine the SOGs of spin Hamiltonians containing Heisenberg exchange, DM interactions, $\Gamma$ exchange, and SIA. An antisymmetric DM interaction selects the direction of a DM vector and reduces the spin-rotation symmetry of the Heisenberg model from $\mathbb{SO}(3)$ to $\mathbb{SO}(2)$. Similarly, a symmetric $\Gamma$ interaction or SIA selects the principal axes of tensors and reduces the full $\mathbb{SO}(3)$ symmetry to the subgroup that leaves the tensor invariant. By extending the local symmetry of individual interaction terms to the full space, we identify 6 possible SOGs for bilinear spin Hamiltonians. We then classify a total of 161 NSGs using quotient-group analysis and the fundamental homomorphism theorem as implemented in Ref. \cite{LITVIN1974538,Litvin1977}. The GSGs of magnetically ordered states are subsequently obtained through spontaneous symmetry breaking of the corresponding parent groups. We find 6 possible SOGs for collinear, coplanar, and noncoplanar magnetic configurations. By deducing their normalizers and the associated quotient groups, we classify 90, 264, 157, 264, and 387 NSGs for the 5 nontrivial SOGs. Finally, we derive all the symmetry-allowed forms of the magnon thermal conductivity and spin conductivity tensors for all GSGs.
\begin{table*}[ptb]
	\caption{SOGs $\mathbb S^{\mathcal S}$ and $\mathbb S'^{\mathcal S}$ for different spin Hamiltonians and magnetic configurations.}
	\begin{tabular}{|c|c|c|c|c|c|c|c|}
		\hline
		Magnetic interactions&$\mathbb S^{\mathcal S}$&$\mathbb S'^{\mathcal S}$&Magnetic configurations\\
		\hline
		&$\mathbb Z_{2}^{\mathcal S}\otimes \mathbb{SO}(3)^{\mathcal S}$&$\mathbb C_{\infty v}^{\mathcal S}$& collinear\\
		$H_{\mathrm J}$&&$\mathbb C_{s}^{\mathcal S}$&coplanar\\
		&&$\mathbbm 1$&noncoplanar\\
		\hline
		&$\mathbb Z_{2}^{\mathcal S}\otimes \mathbb{SO}(2)^{\mathcal S}$&$\mathbb{SO}(2)^{\mathcal S}$&collinear\\
		$H_{\mathrm DM}$&&$\mathbb C_{s}^{\mathcal S}$&collinear/coplanar\\
		&$\mathbb Z_{2}^{\mathcal S}$&$\mathbbm 1^{\mathcal S}$&collinear/coplanar/noncoplanar\\
		\hline
		&$\mathbb Z_{2}^{\mathcal S}\otimes \mathbb{O}(2)^{\mathcal S}$&$\mathbb C_{\infty v}^{\mathcal S}$&collinear\\
		&&$\mathbb C_{s}^{\mathcal S}$&collinear/coplanar\\
		&&$\mathbbm 1^{\mathcal S}$&collinear/coplanar/noncoplanar\\
		$H_{\mathrm \Gamma}/H_{\mathrm{SIA}}$&$\mathbb Z_{2}^{\mathcal S}\otimes \mathbb{D}_{2}^{\mathcal S}$&$\mathbb{C}_{2v}^{\mathcal S}$&collinear\\
		&$\mathbb Z_{2}^{\mathcal S}\otimes \mathbb{C}_{2}^{\mathcal S}$&$\mathbb{C}_{2}^{\mathcal S}$&collinear\\
		&&$\mathbb C_{s}^{\mathcal S}$&collinear/coplanar\\
		&$\mathbb Z_{2}^{\mathcal S}$&$\mathbbm 1^{\mathcal S}$&collinear/coplanar/noncoplanar\\
		\hline
	\end{tabular}
\end{table*}

\textit{Construction of Generalized Spin Groups.--}We start from the spin models to systematically derive all GSGs for spin Hamiltonians and magnetic structures.

\textit{Symmetry of spin Hamiltonians.} The properties of magnons are commonly described in terms of effective spin Hamiltonians \cite{Szilva2023MagneticInteractions}, which generally contain bilinear exchange interactions \cite{DZYALOSHINSKY1958241,PhysRev.120.91}:
\begin{equation}\begin{array}{llll}
	H=H_\mathrm J+H_{\mathrm{DM}}+H_{\Gamma}+H_{\mathrm{SIA}}
	\\=
	\sum_{ij}
	\left[
	J_{ij}\mathbf{S}_i\cdot\mathbf{S}_j
	+
	\mathbf{D}_{ij}\cdot
	\left(\mathbf{S}_i\times\mathbf{S}_j\right)
	+
	\mathbf{S}_i\cdot
	\boldsymbol{\Gamma}_{ij}\cdot
	\mathbf{S}_j
	\right]\\+
	\sum_i
	\mathbf{S}_i\cdot\mathbf{A}_i\cdot\mathbf{S}_i.
	\end{array}
\end{equation}
Here, the first three terms are the isotropic Heisenberg exchange, the antisymmetric DM interaction, and the symmetric $\Gamma$ exchange, respectively, while $\mathbf{A}_i$ is a $3\times3$ real symmetric matrix describing the SIA at site $i$. When studying the properties of a spin model, it is essential to identify the symmetry group that leaves the Hamiltonian invariant. The symmetry group of this Hamiltonian is defined as
\begin{equation}
\mathbb G=\{\{L||R\}|\{L||R\}H\{L||R\}^{-1}=H\}.
\end{equation}
Here $L$ is an operation acting in spin space, $R$ is an operation acting in lattice space. If SOC-induced interactions are absent, namely $\mathbf D_{ij}=\mathbf \Gamma_{ij}=\mathbf A_{i}=0$, the Hamiltonian reduces to the isotropic Heisenberg model $H_{J}$, which commutes with global rotation operator $[H_{J},e^{-i\bm\theta\cdot\mathbf S/\hbar}]=0$, where $\mathbf S=\sum_{i}\mathbf S_{i}$ represents summation of spins. Thus, $H_{J}$ preserves the symmetry of group $\mathbb Z_{2}^{\mathcal S} \otimes \mathbb{SO}(3)^{\mathcal S} \otimes \mathbb C^{\mathcal L}$, as illustrated in Fig. 1(a). Here, group $\mathbb Z_{2}=\{E,T\}$, $\mathbb C$ denotes the crystallographic groups, the group $\mathbb{SO}(3)$ encodes the full rotation symmetry, and the superscripts $\mathcal S$ and $\mathcal L$ specify the spin and the lattice spaces, respectively. When SOC-induced interactions are present, the symmetry group $\mathbb G$ is reduced from the fully decoupled group to its subgroup. Some pure spin operations remain, while other spin operations survive only with an appropriate lattice-space operation, see Fig. 1(b). This group $\mathbb G$ can be expressed as the direct product of (i) a residual SOG $\mathbb S^{\mathcal S}$, which consists of the pure operations acting in the spin space, and (ii) a NSG $\mathbb N^{\mathcal{SL}}$ that contains no pure spin operations other than the identity. Since the spin Hamiltonian itself preserves time-reversal symmetry, the SOG can be written as $\mathbb S^{\mathcal S}=\mathbb Z_{2}^{\mathcal S} \otimes\mathbb M^{\mathcal S}$, where the group $\mathbb M^{\mathcal S}$ is the subgroup of $\mathbb{SO}(3)^{\mathcal S}$. This decomposition separates two distinct effects of SOC-induced interactions. First, anisotropic spin interactions reduce the spin-only symmetry from $\mathbb{SO}(3)^{\mathcal S}$ to a smaller subgroup $\mathbb M^{\mathcal S}$. Second, the lattice symmetry determines which of the remaining or broken spin operations can appear as combined spin-lattice operations in $\mathbb N^{\mathcal{SL}}$. Therefore, the construction of GSGs reduces to determining all possible SOGs $\mathbb M^{\mathcal S}$ and compatible NSGs $\mathbb N^{\mathcal{SL}}$ that leave the spin Hamiltonian invariant.

The SOG $M^S$ is determined by the spin-space operations that leave the Hamiltonian invariant acting on $\mathbf S_{i}$ itself while trivially on sites. It can be obtained by (i) first identifying the local spin symmetry of each interacting site or bond and (ii) then taking the intersection of all local symmetries. We consider the DM interaction as an illustrative example. The local symmetry for a single spin pair consists of all spin rotations that leave the vector $\mathbf{D}$ invariant, corresponding to the group $\mathbb{SO}(2)^{\mathcal S}$, with the rotation axis parallel to vector $\mathbf{D}$. Then, if the orientations of $\mathbf{D}_{ij}$ are uniform across all interacting pairs, the global spin symmetry remains $\mathbb{SO}(2)^{\mathcal S}$. In contrast, if any two $\mathbf{D}_{ij}$ vectors on different bonds are nonparallel, no nontrivial global spin rotation remains, and the global spin symmetry is reduced to the trivial group $\mathbbm{1}^{\mathcal S}$. In summary, the SOGs of the DM interaction can be either $\mathbb{SO}(2)^{\mathcal S}$ or $\mathbbm{1}^{\mathcal S}$. An analogous analysis can be applied to the $\Gamma$ interaction and results are summarized in Table I. More details are shown in section I of Supplementary Material (SM) \cite{supple-material}. SIA is the on-site counterpart of the $\Gamma$ interaction: it is described by a symmetric second-rank tensor and therefore imposes the same type of spin-only symmetry constraints as $\Gamma$-type terms. 

An equivalent but more systematic way to determine the local spin symmetry is to decompose the bilinear spin interactions into irreducible representations of $\mathbb{SO}(3)$ \cite{Brcker1985,Tung1985,Hall2015}. The standard bilinear interactions in Eq. (1) correspond to angular-momentum channels $\ell=0,1,2$, respectively.  The local symmetry of each interaction is then determined by the isotropy class of the corresponding $\mathbb{SO}(3)$ irrep
\cite{RevModPhys.52.617,IHRIG19841,olive2013symmetry,olive2014symmetry}, as summarized in Table S1. The isotropy classes of $\mathbb{SO}(3)$ irreducible representations are conjugate to closed subgroups of $\mathbb{SO}(3)$ \cite{Olive2019}, including
\begin{equation}
	\mathbb O(2), \mathbb{SO}(2), \mathbb D_{n}, \mathbb C_{n}, \mathbb T, \mathbb O, \mathbb I, \mathbbm{1},
\end{equation}
where $\mathbb O(2)$ and $\mathbb{SO}(2)$ denote the full orthogonal and proper rotation groups in two dimensions, respectively. $\mathbb D_n$ and $\mathbb C_n$ are the dihedral and cyclic groups of order $n$, respectively, and $\mathbb T$, $\mathbb O$, and $\mathbb I$ are the tetrahedral, octahedral, and icosahedral groups. $\mathbbm{1}$ denotes the trivial group containing only the trivial element $E$. For the bilinear interactions considered here, the consequences are simple.  The $\ell=0$ scalar Heisenberg term is invariant under the full spin-rotation group $\mathbb{SO}(3)^{\mathcal S}$.  The $\ell=1$ vector DM term has local $\mathbb{SO}(2)^{\mathcal S}$ symmetry about the direction selected by the DM vector.  The $\ell=2$ sector with uniaxial $3z^2-r^2$ component is invariant under $\mathbb O(2)^{\mathcal S}$, whereas the remaining four biaxial quadrupolar components retain only $\mathbb D_2^{\mathcal S}$ symmetry from Table S1.
	\begin{table*}[ptb]
	\caption{SOGs $\mathbb S'^{\mathcal S}$, normalizers $N(\mathbb S'^{\mathcal S})$, quotient groups $N(\mathbb S'^{\mathcal S})/\mathbb S'^{\mathcal S}$, admissible point groups $\mathbb B^{\mathcal S}$ and number of NSGs of ordered states are listed.}
	\begin{tabular}{|c|c|c|c|c|c|c|c|}
		\hline
		$\mathbb S'^{\mathcal S}$&$N(\mathbb S'^{\mathcal S})$&$N(\mathbb S'^{\mathcal S})/\mathbb S'^{\mathcal S}$& $\mathbb B'^{\mathcal S}$&Number\\
		\hline
		$\mathbb C_{\infty v}$&$\mathbb D_{\infty h}$&$\mathbb C_{i}$&$\mathbb C_{1}$, $\mathbb C_{i}$&90\\
		$\mathbb{SO}(2)$&$\mathbb D_{\infty h}$&$\mathbb C_{2h}$&$\mathbb C_{1}$, $\mathbb C_{2}$, $\mathbb C_{i}$, $\mathbb C_{s}$, $\mathbb C_{2h}$&264\\
		$\mathbb C_{s}$&$\mathbb D_{\infty h}$&$\mathbb D_{\infty}$&$\mathbb C_{1}$,$\mathbb C_{2}$,$\mathbb C_{3}$,$\mathbb C_{4}$,$\mathbb C_{6}$, $\mathbb D_{2}$, $\mathbb D_{3}$, $\mathbb D_{4}$, $\mathbb D_{6}$&157\\
		$\mathbb C_{2v}$&$\mathbb D_{4h}$&$\mathbb C_{2h}$&$\mathbb C_{1}$, $\mathbb C_{2}$, $\mathbb C_{i}$, $\mathbb C_{s}$, $\mathbb C_{2h}$&264\\
		$\mathbb C_{2}$&$\mathbb D_{\infty h}$&$\mathbb D_{\infty h}/\{E,C_{2z}\}$&$\mathbb C_{1}$,$\mathbb C_{2}$,$\mathbb C_{3}$,$\mathbb D_{3}$,$\mathbb C_{i}$,$\mathbb C_{s}$,$\mathbb S_{6}$,$\mathbb C_{2h}$,$\mathbb C_{3h}$,$\mathbb D_{3h}$,$\mathbb C_{3v}$, $\mathbb D_{3d}$&387\\
		\hline
	\end{tabular}
\end{table*}

The NSGs $\mathbb N^{\mathcal{SL}}$ can be derived as follows \cite{Litvin1977}. The group operations in $\mathbb N^{\mathcal{SL}}$ consist of two parts: (i) All operations act solely in the lattice space, forming a subgroup $\mathbb E^{\mathcal L}$; (ii) Combined spin-lattice operations that leave the Hamiltonian invariant, denoted as $\{B_{i}||F_{i}\}$, where $B_{i}\in\mathbb B^{\mathcal S}$ and $F_{i}\in\mathbb F^{\mathcal L}$ are nontrivial group elements acting in the spin space and the lattice space, respectively. $\mathbb B^{\mathcal S}$ and $\mathbb F^{\mathcal L}$ denote the sets of group elements of $\mathbb N^{\mathcal{SL}}$ acting purely in the spin and the lattice spaces,
\begin{equation}\begin{array}{cc}
		\mathbb B^{\mathcal S}=E+\sum_{i}B_{i},\\
		\mathbb F^{\mathcal L}=\mathbb E^{\mathcal L}+\sum_{i}F_{i}\mathbb E^{\mathcal L},
	\end{array}
\end{equation}
respectively. Thus, NSGs can be expressed as
\begin{equation}
	N^{\mathcal{SL}}=\mathbb E^{\mathcal L}+\sum\nolimits_i\{B_{i}||F_{i}\}\mathbb E^{\mathcal L}.
\end{equation}
This group should be a subgroup of the direct product group $\mathbb B^{\mathcal S}\otimes\mathbb F^{\mathcal L}$.
The compatibility between lattice and spin operations is encoded by a homomorphism between two groups $\mathbb F^{\mathcal L}$ and $\mathbb B^{\mathcal S}$, $\rho: \mathbb E^{\mathcal L}\mapsto E, \quad F_i\mathbb E^{\mathcal L} \mapsto B_i$. Since the group $\mathbb E^{\mathcal L}$ is the kernel of homomorphism that maps to the trivial element, it is a normal subgroup of $\mathbb F^{\mathcal L}$ \cite{dummit2004abstract,gallian2021contemporary}. The fundamental theorem of homomorphisms \cite{dummit2004abstract,gallian2021contemporary} gives
\begin{equation}\mathbb F^{\mathcal L}/\mathbb E^{\mathcal L}\cong \mathbb B^{\mathcal S}.
\end{equation}	
Therefore, the construction of the NSGs $N^{\mathcal{SL}}$ proceeds in two steps: (i) Enumerate all normal subgroups $\mathbb E^{\mathcal L}$ of group $\mathbb F^{\mathcal L}$; (ii) Construct all $\mathbb B^{\mathcal S}$ of equivalent conjugate classes by $\mathbb F^{\mathcal L}/\mathbb E^{\mathcal L}$. Note that two groups defined to belong to the same type if they are conjugate subgroups of the direct product of $\mathbb K\times\mathbb A$, where $\mathbb K$ is the general linear group $\mathbb{GL}(3)$ in spin space, and $\mathbb A$ is the affine group $\mathbb{GIL}(3)$ in physical space \cite{LITVIN1974538,Litvin1977}. Following these two procedures, we obtain a total of 161 nontrivial spin point groups, which are summarized in Table S2 of the SM \cite{supple-material}. Here, the rotation groups and reflection groups belong to equivalent conjugate classes due to the presence of time-reversal symmetry in spin space. 

\textit{Symmetry of magnetic configurations.} In contrast to the Hamiltonian, an ordered magnetic state generally breaks time-reversal symmetry, so that time reversal is no longer an independent symmetry operation. The symmetry of an ordered magnetic state can be obtained from the spontaneous symmetry breaking of a high-symmetry Hamiltonian \cite{Toledano1987LandauTheory,RevModPhys.52.617}. The GSG of a magnetic structure, denoted by $\mathbb G'$, can still be decomposed into a direct product of a SOG $\mathbb S'^{\mathcal S}$ and a NSG $\mathbb N'^{\mathcal{SL}}$.

The SOG $\mathbb S'^{\mathcal S}$ depends on three different magnetic configurations: collinear, coplanar, and noncoplanar, therefore $\mathbb S'^{\mathcal S}$ can be obtained by intersecting the symmetry allowed by the magnetic configuration with the spin-only symmetry of the Hamiltonian $\mathbb S^{\mathcal S}$. We first consider the isotropic Heisenberg model without SOC. In this case, the Hamiltonian has the spin-only symmetry $\mathbb Z_{2}^{\mathcal S}\otimes \mathbb{SO}(3)^{\mathcal S}$, which is isomorphic to $\mathbb O(3)^{\mathcal S}$. A collinear magnetic state preserves rotations about the spin axis and a spin-space reflection operation, giving $\mathbb C_{\infty v}^{\mathcal S}=\mathbb{SO}(2)^{\mathcal S} \rtimes \mathbb C_{s}^{\mathcal S}$, see Table I. Also, a coplanar magnetic state preserves only a spin-space reflection operation $\mathbb C_{s}^{\mathcal S}$. A generic noncoplanar state has no nontrivial spin-only symmetry $\mathbbm1^{\mathcal S}$. Here, $\mathbb C_{s}^{\mathcal S}=\{E,C_{2}T\}$ is the order-2 reflection group in spin space. When SOC-induced interactions are present, for example, if all DM vectors are parallel, the spin Hamiltonian has $\mathbb Z_{2}^{\mathcal S}\otimes \mathbb{SO}(2)^{\mathcal S}$ spin-only symmetry. A collinear magnetic state aligned with the DM axis preserves $\mathbb{SO}(2)^{\mathcal S}$ while the reflection operation $\mathbb C_{s}^{\mathcal S}$ is broken, see Table I. A collinear magnetic order perpendicular to the DM axis breaks the $\mathbb{SO}(2)^{\mathcal S}$ rotation symmetry but can preserve the reflection operation $\mathbb C_{s}^{\mathcal S}$. A coplanar magnetic order whose spin plane is perpendicular to DM axis can retain $\mathbb C_{s}^{\mathcal S}$. Finally, in the presence of DM interaction, the SOGs for magnetic ground states are only $\mathbb{SO}(2)^{\mathcal S}$, $\mathbb C_{s}^{\mathcal S}$, or $\mathbbm 1^{\mathcal S}$ as shown in Table I. The corresponding SOGs for magnetic states with the $\Gamma$ interaction or SIA are also listed in Table I. In total, 6 distinct SOGs arise for magnetic structures. Further details are shown in Sec. II of SM \cite{supple-material}.

Among these 6 SOGs, the trivial group \(\mathbbm 1^{\mathcal S}\) can describe noncoplanar magnetic structures. Their nontrivial symmetries can be described by the 598 inequivalent NSGs classified by Litvin~\cite{Litvin1977}. Analogous to obtaining type-I and type-III magnetic point groups from type-II magnetic point groups, these ordered-state NSGs can also be constructed from the 161 NSGs we derived above through breaking of the time-reversal symmetry. 

For magnetic structures with other 5 SOGs, the allowed NSGs are determined by a normalizer-quotient construction. Given an SOG $\mathbb S'^{\mathcal S}$, we first deduce its normalizer $N\left(\mathbb S'^{\mathcal S}\right)$ in the full orthogonal group $\mathbb O(3)^{\mathcal S}$ to identify all spin-space operations that are compatible with the given spin-only symmetry. The quotient $N\left(\mathbb S'^{\mathcal S}\right)/\mathbb S'^{\mathcal S}$ then gives the effective residual spin-space symmetry, acting faithfully after the spin-only operations have been factored out. We enumerate the crystallographically admissible subgroups of this quotient group and use them as the spin part $\mathbb B'^{\mathcal S}$ of the NSGs. Finally, combining these allowed $\mathbb B'^{\mathcal S}$ with the corresponding lattice point-group operations yields the full set of NSGs. This construction can be illustrated by the collinear case. The SOG of a collinear state without SOC is given by $\mathbb C_{\infty v}^{\mathcal S}$. Its normalizer in $\mathbb O(3)^{\mathcal S}$ is $N\left(\mathbb C_{\infty v}^{\mathcal S}\right)=\mathbb D_{\infty h}^{\mathcal S}$, and the quotient group is $\mathbb D_{\infty h}^{\mathcal S}/\mathbb C_{\infty v}^{\mathcal S}\simeq \mathbb C_i$. The two subgroups of $\mathbb C_i$, namely $\mathbb C_1$ and $\mathbb C_i$, generate a total of 90 NSGs, reproducing the spin-group classification of collinear magnets without SOC~\cite{PhysRevX.12.021016}. When the DM interaction is included and the magnetic moments are assumed to be parallel to the DM vector, the SOG is reduced from $\mathbb C_{\infty v}^{\mathcal S}$ to $\mathbb{SO}(2)^{\mathcal S}$. In this case, $N\left[\mathbb{SO}(2)^{\mathcal S}\right]=\mathbb D_{\infty h}^{\mathcal S}$ and $\mathbb D_{\infty h}^{\mathcal S}/\mathbb{SO}(2)^{\mathcal S}\simeq \mathbb C_{2h}$. The crystallographically admissible subgroups are $\mathbb C_1$, $\mathbb C_2$, $\mathbb C_i$, $\mathbb C_s$, and $\mathbb C_{2h}$. Using these groups as the allowed spin part $\mathbb B'^{\mathcal S}$, we obtain a total of 264 NSGs. The results for the 5 nontrivial SOGs are summarized in Table II, and the details are provided in Table S3-S7 in the SM \cite{supple-material}.

\textit{Applications to Magnon Transport.--}
Altermagnetism has attracted rapidly growing interest as a compensated magnetic phase hosting nonrelativistic spin-split bands \cite{PhysRevX.12.031042,PhysRevX.12.040501}, whose symmetry is naturally captured by spin groups. This framework has also been applied to nonrelativistic magnon spin Seebeck and spin Nernst responses \cite{PhysRevB.108.L180401,PhysRevB.110.094427,Wu2025MagnonSplitting,g5xq-z15c}. More recently, however, some studies have revealed relativistic magnon transport phenomena in altermagnets \cite{PhysRevB.111.L020412,hm4g-rl3y}. These SOC-dependent responses lie beyond a purely nonrelativistic spin-group description. Conventionally, they are analyzed using magnetic-group symmetry, which treats spin and lattice transformations as locked by SOC. However, this full spin-lattice locking lowers the symmetry and admits a broad set of transport channels, which may remain forbidden by residual symmetries. The GSG framework bridges these limits by retaining residual spin-only and partially locked spin-lattice symmetries in the presence of SOC.

Here, we firstly derive the complete constraints imposed by the six SOGs and 598 NSGs on the thermal- and spin-conductivity tensors, which are summarized in Sec.~III of the SM \cite{supple-material}. We then use MnPS$_3$ as a representative example to compare the predictions of magnetic group, spin group, and GSG for the magnon spin Nernst effect. MnPS$_3$ is a layered van der Waals insulator with a monoclinic crystal structure \cite{Ouvrard1985}. Below the Néel temperature $T_N \approx 78$ K, the Mn$^{2+}$ moments form antiferromagnetic order within $ab$ plane and are coupled ferromagnetically between adjacent layers \cite{Wildes1998,Nawwar2025}, resulting in the magnetic point group $2'/m$. This magnetic point group allows five magnon spin Nernst channels $\sigma_{xy}^{z}$, $\sigma_{yz}^{z}$, $\sigma_{xy}^{x}$, $\sigma_{yz}^{x}$, and $\sigma_{xz}^{y}$. We next consider the $J_1$-$D_2$ honeycomb-antiferromagnet model proposed for MnPS$_3$ \cite{PhysRevLett.117.217202} (see Eq.~S10 in the SM). Its GSG comprises a spin-only $\mathbb{SO}(2)^{\mathcal S}$ symmetry, corresponding to rotations about the DM-vector direction, and a residual spin-lattice group ${}^m2/{}^{2}m$. The spin-only $\mathbb{SO}(2)^{\mathcal S}$ symmetry restricts the spin polarization to the $z$ direction, thereby forbidding channels $\sigma_{xy}^{x}$, $\sigma_{yz}^{x}$, and $\sigma_{xz}^{y}$, and thus allow channels $\sigma_{xy}^{z}$ and $\sigma_{yz}^{z}$. In the nonrelativistic limit $D_2=0$, the spin group symmetry is $\mathbb C_{\infty v}^{\mathcal S}\times{}^{1}2/^{\bar1}m$ in the case, and it instead only permits one channel, $\sigma_{xz}^{z}$. This prediction is not in agreement with the theoretical predictions and experimental evidence for MnPS$_3$ \cite{PhysRevLett.117.217202,PhysRevLett.117.217203,PhysRevB.96.134425}.	

\textit{Conclusions.--}In conclusion, we have introduced GSGs as a unified symmetry framework for magnons with or without SOC-induced spin interactions. This framework goes beyond both conventional magnetic groups, which assume complete spin-lattice locking, and spin groups, which assume separable spin and lattice spaces. Starting from the bilinear spin Hamiltonians, we classified the GSGs for Heisenberg, DM, $\Gamma$ interactions, and SIA. We found 6 possible SOGs and 161 NSGs. The GSGs of ordered magnetic states were then obtained by spontaneous symmetry breaking of the corresponding parent groups. The resulting SOGs fall into 6 classes, which generate 90, 264, 157, 264, 387, and 598 NSGs, respectively. Finally, we apply GSGs to magnon transport and derive all the symmetry-allowed magnon thermal conductivity and spin conductivity tensors. The MnPS$_3$ example show that conventional magnetic groups can allow transverse magnon response channels forbidden by residual symmetries of GSGs, while spin groups cannot account for responses induced by SOC. GSGs thus provide a unified symmetry framework for analyzing magnon properties with or without SOC.

\textit{Acknowledgments--}This work was financially supported by the National Natural Science Foundation of China (Grants No. 12188101, No. 12334007, No. 12474233, and No.1260042396), the National Key R\&D Program of China (No. 2025YFA1411301, No. 2025YFA1411302, and No. 2022YFA1403601), Innovation Program for Quantum Science and Technology (Grant No. 2021ZD0301902, and No. 2024ZD0300101), and Natural Science Foundation of Jiangsu Province (No. BK20233001, and No. BK20243011), Fundamental and Interdisciplinary Disciplines Breakthrough Plan of the Ministry of Education of China (JYB2025XDXM411), the Fundamental Research Funds for the Central Universities (No. KG202501). This work has been supported by the New Cornerstone Science Foundation.

\bibliography{bibl}

@article{PhysRevB.111.L020412,
	title   = {Spontaneous crystal thermal Hall effect in insulating altermagnets},
	author  = {Hoyer, Rhea and Jaeschke-Ubiergo, Rodrigo and Ahn, Kyo-Hoon
	and {\v{S}}mejkal, Libor and Mook, Alexander},
	journal = {Physical Review B},
	volume  = {111},
	number  = {2},
	pages   = {L020412},
	year    = {2025},
	doi     = {10.1103/PhysRevB.111.L020412}
}

@article{hm4g-rl3y,
	title = {Spin-orbit coupling driven handedness switching of magnons in altermagnets},
	author = {Li, Wen-Tong and Wu, Yu-Biao and Zhuang, Lin and Wang, Jian-Tao and Liu, Wu-Ming},
	journal = {Phys. Rev. B},
	volume = {113},
	issue = {22},
	pages = {224403},
	numpages = {12},
	year = {2026},
	month = {Jun},
	publisher = {American Physical Society},
	doi = {10.1103/hm4g-rl3y},
	url = {https://link.aps.org/doi/10.1103/hm4g-rl3y}
}

@article{PhysRevX.12.040501,
	title   = {Emerging Research Landscape of Altermagnetism},
	author  = {{\v{S}}mejkal, Libor and Sinova, Jairo and Jungwirth, Tomas},
	journal = {Physical Review X},
	volume  = {12},
	number  = {4},
	pages   = {040501},
	year    = {2022},
	doi     = {10.1103/PhysRevX.12.040501}
}

@article{PhysRevX.14.031038,
	title = {Enumeration and Representation Theory of Spin Space Groups},
	author = {Chen, Xiaobing and Ren, Jun and Zhu, Yanzhou and Yu, Yutong and Zhang, Ao and Liu, Pengfei and Li, Jiayu and Liu, Yuntian and Li, Caiheng and Liu, Qihang},
	journal = {Phys. Rev. X},
	volume = {14},
	issue = {3},
	pages = {031038},
	numpages = {33},
	year = {2024},
	month = {Aug},
	publisher = {American Physical Society},
	doi = {10.1103/PhysRevX.14.031038},
	url = {https://link.aps.org/doi/10.1103/PhysRevX.14.031038}
}

@article{PhysRevB.110.094427,
	title     = {Atomistic Spin Dynamics Simulations of Magnonic Spin Seebeck and Spin Nernst Effects in Altermagnets},
	author    = {Wei{\ss}enhofer, Markus and Marmodoro, Alberto},
	journal   = {Phys. Rev. B},
	volume    = {110},
	issue     = {9},
	pages     = {094427},
	numpages  = {14},
	year      = {2024},
	month     = {Sep},
	publisher = {American Physical Society},
	doi       = {10.1103/PhysRevB.110.094427},
	url       = {https://doi.org/10.1103/PhysRevB.110.094427}
}

@article{Wu2025MagnonSplitting,
	title   = {Magnon Splitting and Magnon Spin Transport in Altermagnets},
	author  = {Wu, Kun and Dong, Jianting and Zhu, Meng and Zheng, Fanxing and Zhang, Jia},
	journal = {Chinese Physics Letters},
	volume  = {42},
	number  = {7},
	pages   = {070702},
	year    = {2025},
	doi     = {10.1088/0256-307X/42/7/070702},
	url     = {https://doi.org/10.1088/0256-307X/42/7/070702}
}

@article{g5xq-z15c,
	title     = {Altermagnet-Driven Magnon Spin Splitting Nernst Effect},
	author    = {Yang, Yuben and Wang, Di and Yang, Bin and Wang, Peng
	and Mu, Yuxuan and Tian, Yuanzhe and Zheng, Bowen
	and Qin, Weijie and Wang, Kaiyuan and Huang, Biying
	and Wang, Baigeng and Wan, Xiangang and Wu, Di},
	journal   = {Phys. Rev. Lett.},
	volume    = {136},
	issue     = {2},
	pages     = {026701},
	numpages  = {7},
	year      = {2026},
	month     = {Jan},
	publisher = {American Physical Society},
	doi       = {10.1103/g5xq-z15c},
	url       = {https://doi.org/10.1103/g5xq-z15c}
}

@article{PhysRevB.92.155138,
	title     = {Symmetry-Imposed Shape of Linear Response Tensors},
	author    = {Seemann, M. and K{\"o}dderitzsch, D. and Wimmer, S. and Ebert, H.},
	journal   = {Phys. Rev. B},
	volume    = {92},
	issue     = {15},
	pages     = {155138},
	numpages  = {12},
	year      = {2015},
	month     = {Oct},
	publisher = {American Physical Society},
	doi       = {10.1103/PhysRevB.92.155138},
	url       = {https://doi.org/10.1103/PhysRevB.92.155138}
}

@article{PhysRevB.108.085140,
	title = {First-principles study of the spin-orbit coupling contribution to anisotropic magnetic interactions},
	author = {Wang, Di and Bo, Xiangyan and Tang, Feng and Wan, Xiangang},
	journal = {Phys. Rev. B},
	volume = {108},
	issue = {8},
	pages = {085140},
	numpages = {11},
	year = {2023},
	month = {Aug},
	publisher = {American Physical Society},
	doi = {10.1103/PhysRevB.108.085140},
	url = {https://link.aps.org/doi/10.1103/PhysRevB.108.085140}
}

@article{PhysRevB.108.L180401,
	title     = {Efficient Spin Seebeck and Spin Nernst Effects of Magnons in Altermagnets},
	author    = {Cui, Qirui and Zeng, Bowen and Cui, Ping and Yu, Tao and Yang, Hongxin},
	journal   = {Phys. Rev. B},
	volume    = {108},
	issue     = {18},
	pages     = {L180401},
	numpages  = {7},
	year      = {2023},
	month     = {Nov},
	publisher = {American Physical Society},
	doi       = {10.1103/PhysRevB.108.L180401},
	url       = {https://doi.org/10.1103/PhysRevB.108.L180401}
}

@article{PhysRevX.14.031039,
	title = {Enumeration of Spin-Space Groups: Toward a Complete Description of Symmetries of Magnetic Orders},
	author = {Jiang, Yi and Song, Ziyin and Zhu, Tiannian and Fang, Zhong and Weng, Hongming and Liu, Zheng-Xin and Yang, Jian and Fang, Chen},
	journal = {Phys. Rev. X},
	volume = {14},
	issue = {3},
	pages = {031039},
	numpages = {25},
	year = {2024},
	month = {Aug},
	publisher = {American Physical Society},
	doi = {10.1103/PhysRevX.14.031039},
	url = {https://link.aps.org/doi/10.1103/PhysRevX.14.031039}
}

@book{White2007,
	author    = {White, Robert M.},
	title     = {Quantum Theory of Magnetism: Magnetic Properties of Materials},
	edition   = {3},
	series    = {Springer Series in Solid-State Sciences},
	volume    = {32},
	publisher = {Springer},
	address   = {Berlin},
	year      = {2007},
	doi       = {10.1007/978-3-540-69025-2}
}

@article{PhysRevLett.117.217202,
	title = {Spin Nernst Effect of Magnons in Collinear Antiferromagnets},
	author = {Cheng, Ran and Okamoto, Satoshi and Xiao, Di},
	journal = {Phys. Rev. Lett.},
	volume = {117},
	issue = {21},
	pages = {217202},
	numpages = {5},
	year = {2016},
	month = {Nov},
	publisher = {American Physical Society},
	doi = {10.1103/PhysRevLett.117.217202},
	url = {https://link.aps.org/doi/10.1103/PhysRevLett.117.217202}
}

@article{PhysRevLett.117.217203,
	title = {Magnon Spin Nernst Effect in Antiferromagnets},
	author = {Zyuzin, Vladimir A. and Kovalev, Alexey A.},
	journal = {Phys. Rev. Lett.},
	volume = {117},
	issue = {21},
	pages = {217203},
	numpages = {5},
	year = {2016},
	month = {Nov},
	publisher = {American Physical Society},
	doi = {10.1103/PhysRevLett.117.217203},
	url = {https://link.aps.org/doi/10.1103/PhysRevLett.117.217203}
}

@article{PhysRevB.96.134425,
	title = {Experimental evidence consistent with a magnon Nernst effect in the antiferromagnetic insulator ${\mathrm{MnPS}}_{3}$},
	author = {Shiomi, Y. and Takashima, R. and Saitoh, E.},
	journal = {Phys. Rev. B},
	volume = {96},
	issue = {13},
	pages = {134425},
	numpages = {9},
	year = {2017},
	month = {Oct},
	publisher = {American Physical Society},
	doi = {10.1103/PhysRevB.96.134425},
	url = {https://link.aps.org/doi/10.1103/PhysRevB.96.134425}
}

@article{McClarty2018TopologicalMagnons,
	author  = {McClarty, P. A. and Dong, X.-Y. and Gohlke, M.
	and Rau, J. G. and Pollmann, F. and Moessner, R.
	and Penc, K.},
	title   = {Topological Magnons in Kitaev Magnets at High Fields},
	journal = {Phys. Rev. B},
	volume  = {98},
	pages   = {060404(R)},
	year    = {2018},
	doi     = {10.1103/PhysRevB.98.060404}
}

@article{Liu2022SIA,
	author  = {Liu, Xiaoyu and Churchill, Derek and Kee, Hae-Young},
	title   = {Theoretical Analysis of Single-Ion Anisotropy in
	{$d^3$} Mott Insulators},
	journal = {Phys. Rev. B},
	volume  = {106},
	pages   = {035122},
	year    = {2022},
	doi     = {10.1103/PhysRevB.106.035122}
}

@article{Szilva2023MagneticInteractions,
	author  = {Szilva, Attila and Kvashnin, Yaroslav and Stepanov, Evgeny A.
	and Nordstr{\"o}m, Lars and Eriksson, Olle
	and Lichtenstein, Alexander I. and Katsnelson, Mikhail I.},
	title   = {Quantitative Theory of Magnetic Interactions in Solids},
	journal = {Rev. Mod. Phys.},
	volume  = {95},
	pages   = {035004},
	year    = {2023},
	doi     = {10.1103/RevModPhys.95.035004}
}

@article{Onose2010MagnonHall,
	title = {Observation of the Magnon Hall Effect},
	author = {Onose, Y. and Ideue, T. and Katsura, H. and Shiomi, Y. and Nagaosa, N. and Tokura, Y.},
	journal = {Science},
	volume = {329},
	pages = {297--299},
	year = {2010},
	doi = {10.1126/science.1188260}
}

@article{Chen2025,
	title = {Unconventional magnons in collinear magnets dictated by spin space groups},
	volume = {640},
	ISSN = {1476-4687},
	url = {http://dx.doi.org/10.1038/s41586-025-08715-7},
	DOI = {10.1038/s41586-025-08715-7},
	number = {8058},
	journal = {Nature},
	publisher = {Springer Science and Business Media LLC},
	author = {Chen,  Xiaobing and Liu,  Yuntian and Liu,  Pengfei and Yu,  Yutong and Ren,  Jun and Li,  Jiayu and Zhang,  Ao and Liu,  Qihang},
	year = {2025},
	month = Mar,
	pages = {349-354}
}

@article{Laurell2018MagnonThermalHall,
	title = {Magnon thermal Hall effect in kagome antiferromagnets with Dzyaloshinskii-Moriya interactions},
	author = {Laurell, Pontus and Fiete, Gregory A.},
	journal = {Physical Review B},
	volume = {98},
	number = {9},
	pages = {094419},
	year = {2018},
	publisher = {American Physical Society},
	doi = {10.1103/PhysRevB.98.094419}
}

@article{Matsumoto2020NonreciprocalMagnons,
	title = {Nonreciprocal magnons due to symmetric anisotropic exchange interaction in honeycomb antiferromagnets},
	author = {Matsumoto, Takuya and Hayami, Satoru},
	journal = {Physical Review B},
	volume = {101},
	number = {22},
	pages = {224419},
	year = {2020},
	publisher = {American Physical Society},
	doi = {10.1103/PhysRevB.101.224419}
}

@article{Scheie2023NiPS3,
	title = {Spin wave Hamiltonian and anomalous scattering in ${\rm NiPS}_3$},
	author = {Scheie, A. and Park, Pyeongjae and Villanova, J. W. and
	Granroth, G. E. and Sarkis, C. L. and Zhang, Hao and
	Stone, M. B. and Park, Je-Geun and Okamoto, S. and
	Berlijn, T. and Tennant, D. A.},
	journal = {Physical Review B},
	volume = {108},
	number = {10},
	pages = {104402},
	year = {2023},
	publisher = {American Physical Society},
	doi = {10.1103/PhysRevB.108.104402}
}

@article{Dimmock1962MagneticCrystals,
	title   = {Symmetry Properties of Wave Functions in Magnetic Crystals},
	author  = {Dimmock, J. O. and Wheeler, R. G.},
	journal = {Physical Review},
	volume  = {127},
	pages   = {391--404},
	year    = {1962},
	doi     = {10.1103/PhysRev.127.391}
}

@article{Harris1963YIG,
	title   = {Spin-Wave Spectra of Yttrium and Gadolinium Iron Garnet},
	author  = {Harris, A. B.},
	journal = {Physical Review},
	volume  = {132},
	pages   = {2398--2417},
	year    = {1963},
	doi     = {10.1103/PhysRev.132.2398}
}

@article{Corticelli2022SpinSpaceMagnonTopology,
	title   = {Spin-space groups and magnon band topology},
	author  = {Corticelli, Alberto and Moessner, Roderich and McClarty, Paul A.},
	journal = {Physical Review B},
	volume  = {105},
	pages   = {064430},
	year    = {2022},
	doi     = {10.1103/PhysRevB.105.064430}
}

@article{Mook2014TopologicalMagnonInsulator,
	title   = {Edge States in Topological Magnon Insulators},
	author  = {Mook, Alexander and Henk, J{\"u}rgen and Mertig, Ingrid},
	journal = {Physical Review B},
	volume  = {90},
	pages   = {024412},
	year    = {2014},
	doi     = {10.1103/PhysRevB.90.024412}
}

@article{Katsura2010ThermalHall,
	title = {Theory of the Thermal Hall Effect in Quantum Magnets},
	author = {Katsura, Hosho and Nagaosa, Naoto and Lee, Patrick A.},
	journal = {Physical Review Letters},
	volume = {104},
	pages = {066403},
	year = {2010},
	doi = {10.1103/PhysRevLett.104.066403}
}

@article{Matsumoto2011RotatingMagnon,
	title = {Theoretical Prediction of a Rotating Magnon Wave Packet in Ferromagnets},
	author = {Matsumoto, Ryo and Murakami, Shuichi},
	journal = {Physical Review Letters},
	volume = {106},
	pages = {197202},
	year = {2011},
	doi = {10.1103/PhysRevLett.106.197202}
}

@article{Olive2019,
	title = {Effective computation of SO(3) and O(3) linear
	representation symmetry classes},
	volume = {7},
	ISSN = {2326-7186},
	number = {3},
	journal = {Mathematics and Mechanics of Complex Systems},
	publisher = {Mathematical Sciences Publishers},
	author = {Olive,  Marc},
	year = {2019},
	month = sept,
	pages = {203-237}
}

@article{Ouvrard1985,
	title = {Structural determination of some MPS3 layered phases (M = Mn,  Fe,  Co,  Ni and Cd)},
	volume = {20},
	ISSN = {0025-5408},
	url = {http://dx.doi.org/10.1016/0025-5408(85)90092-3},
	DOI = {10.1016/0025-5408(85)90092-3},
	number = {10},
	journal = {Materials Research Bulletin},
	publisher = {Elsevier BV},
	author = {Ouvrard,  G. and Brec,  R. and Rouxel,  J.},
	year = {1985},
	month = oct,
	pages = {1181-1189}
}

@article{Wildes1998,
	title = {Spin waves and the critical behaviour of the magnetization in},
	volume = {10},
	ISSN = {1361-648X},
	url = {http://dx.doi.org/10.1088/0953-8984/10/28/020},
	DOI = {10.1088/0953-8984/10/28/020},
	number = {28},
	journal = {Journal of Physics: Condensed Matter},
	publisher = {IOP Publishing},
	author = {Wildes,  A R and Roessli,  B and Lebech,  B and Godfrey,  K W},
	year = {1998},
	month = {july},
	pages = {6417-6428}
}

@article{Nawwar2025,
	title = {Large thermal Hall effect in MnPS3},
	volume = {88},
	ISSN = {1361-6633},
	url = {http://dx.doi.org/10.1088/1361-6633/adf916},
	DOI = {10.1088/1361-6633/adf916},
	number = {8},
	journal = {Reports on Progress in Physics},
	publisher = {IOP Publishing},
	author = {Nawwar,  Mohamed and Neumann,  Robin R and Wen,  Jiamin and Mertig,  Ingrid and Mook,  Alexander and Heremans,  Joseph P},
	year = {2025},
	month = aug,
	pages = {080503}
}

@article{LITVIN1974538,
	title = {Spin groups},
	journal = {Physica},
	volume = {76},
	number = {3},
	pages = {538-554},
	year = {1974},
	issn = {0031-8914},
	author = {D.B. Litvin and W. Opechowski},
}

@article{PhysRevX.14.031037,
	title = {Spin Space Groups: Full Classification and Applications},
	author = {Xiao, Zhenyu and Zhao, Jianzhou and Li, Yanqi and Shindou, Ryuichi and Song, Zhi-Da},
	journal = {Phys. Rev. X},
	volume = {14},
	issue = {3},
	pages = {031037},
	numpages = {33},
	year = {2024},
	month = {Aug},
	publisher = {American Physical Society},
	doi = {10.1103/PhysRevX.14.031037},
	url = {https://link.aps.org/doi/10.1103/PhysRevX.14.031037}
}

@article{PhysRevX.15.031006,
	title = {Multipolar Anisotropy in Anomalous Hall Effect from Spin-Group Symmetry Breaking},
	author = {Liu, Zheng and Wei, Mengjie and Peng, Wenzhi and Hou, Dazhi and Gao, Yang and Niu, Qian},
	journal = {Phys. Rev. X},
	volume = {15},
	issue = {3},
	pages = {031006},
	numpages = {23},
	year = {2025},
	month = {Jul},
	publisher = {American Physical Society},
	doi = {10.1103/PhysRevX.15.031006},
	url = {https://link.aps.org/doi/10.1103/PhysRevX.15.031006}
}

@book{Toledano1987LandauTheory,
	author = {Toledano, Pierre and Toledano, Jean-Claude},
	title = {The Landau Theory of Phase Transitions: Application to Structural, Incommensurate, Magnetic and Liquid Crystal Systems},
	publisher = {World Scientific},
	year = {1987}
}

@book{Shubnikov1964Colored,
	author    = {Shubnikov, A. V. and Belov, N. V.},
	title     = {Colored Symmetry},
	publisher = {Pergamon Press},
	address   = {Oxford},
	year      = {1964}
}

@incollection{Opechowski1965Magnetic,
	author    = {Opechowski, W. and Guccione, R.},
	title     = {Magnetic Symmetry},
	booktitle = {Magnetism},
	editor    = {Rado, G. T. and Suhl, H.},
	volume    = {2A},
	pages     = {105--165},
	publisher = {Academic Press},
	address   = {New York},
	year      = {1965}
}

@book{Birss1964Symmetry,
	author    = {Birss, R. R.},
	title     = {Symmetry and Magnetism},
	publisher = {North-Holland},
	address   = {Amsterdam},
	year      = {1964}
}

@article{Bradley1968Magnetic,
	author  = {Bradley, C. J. and Davies, B. L.},
	title   = {Magnetic Groups and Their Corepresentations},
	journal = {Rev. Mod. Phys.},
	volume  = {40},
	pages   = {359--379},
	year    = {1968},
	doi     = {10.1103/RevModPhys.40.359}
}

@article{Kleiner1966Transport,
	author  = {Kleiner, W. H.},
	title   = {Space-Time Symmetry of Transport Coefficients},
	journal = {Phys. Rev.},
	volume  = {142},
	pages   = {318},
	year    = {1966},
	doi     = {10.1103/PhysRev.142.318}
}

@article{PhysRevX.12.021016,
	title = {Spin-Group Symmetry in Magnetic Materials with Negligible Spin-Orbit Coupling},
	author = {Liu, Pengfei and Li, Jiayu and Han, Jingzhi and Wan, {Xiangang} and Liu, Qihang},
	journal = {Phys. Rev. X},
	volume = {12},
	issue = {2},
	pages = {021016},
	numpages = {19},
	year = {2022},
	month = {Apr},
	publisher = {American Physical Society},
	doi = {10.1103/PhysRevX.12.021016},
	url = {https://link.aps.org/doi/10.1103/PhysRevX.12.021016}
}

@misc{supple-material,
	note={See Supplemental Material for symmetry of spin Hamiltonians, symmetry of magnetic configurations, and magnon thermal and spin conductivity tensors.}
}

@article{Litvin1977,
	title = {Spin point groups},
	volume = {33},
	ISSN = {0567-7394},
	url = {http://dx.doi.org/10.1107/S0567739477000709},
	DOI = {10.1107/s0567739477000709},
	number = {2},
	journal = {Acta Crystallographica Section A},
	publisher = {International Union of Crystallography (IUCr)},
	author = {Litvin,  D. B.},
	year = {1977},
	month = mar,
	pages = {279-287}
}

@article{PhysRevX.12.031042,
	title = {Beyond Conventional Ferromagnetism and Antiferromagnetism: A Phase with Nonrelativistic Spin and Crystal Rotation Symmetry},
	author = {\ifmmode \check{S}\else \v{S}\fi{}mejkal, Libor and Sinova, Jairo and Jungwirth, Tomas},
	journal = {Phys. Rev. X},
	volume = {12},
	issue = {3},
	pages = {031042},
	numpages = {16},
	year = {2022},
	month = {Sep},
	publisher = {American Physical Society},
}

@article{1qd6-ddzl,
title = {Revealing spin and spatial symmetry decoupling in systems with Dzyaloshinskii-Moriya interaction},
author = {Mu, Yuxuan and Wang, Di and Wan, Xiangang},
journal = {Phys. Rev. B},
volume = {113},
issue = {1},
pages = {014432},
numpages = {7},
year = {2026},
month = {Jan},
publisher = {American Physical Society},
doi = {10.1103/1qd6-ddzl},
url = {https://link.aps.org/doi/10.1103/1qd6-ddzl}
}

@article{Etxebarria:cam5007,
author = "Etxebarria, Jesus and Perez-Mato, J. Manuel and Tasci, Emre S. and Elcoro, Luis",
title = "{Crystal tensor properties of magnetic materials with and without spin{--}orbit coupling. Application of spin point groups as approximate symmetries}",
journal = "Acta Crystallographica Section A",
year = "2025",
volume = "81",
number = "4",
pages = "317--338",
month = "Jul",
}

@article{RevModPhys.52.617,
	title = {Symmetry defects and broken symmetry. Configurations Hidden Symmetry},
	author = {Michel, Louis},
	journal = {Rev. Mod. Phys.},
	volume = {52},
	issue = {3},
	pages = {617--651},
	numpages = {0},
	year = {1980},
	month = {Jul},
	publisher = {American Physical Society},
	doi = {10.1103/RevModPhys.52.617},
	url = {https://link.aps.org/doi/10.1103/RevModPhys.52.617}
}

@article{IHRIG19841,
	title = {Pattern selection with O(3) symmetry},
	journal = {Physica D: Nonlinear Phenomena},
	volume = {13},
	number = {1},
	pages = {1-33},
	year = {1984},
	issn = {0167-2789},
	author = {E. Ihrig and M. Golubitsky},
}

@article{olive2014symmetry,
	title={Symmetry classes for odd-order tensors},
	author={Olive, Marc and Auffray, Nicolas},
	journal={ZAMM-Journal of Applied Mathematics and Mechanics/Zeitschrift f{\"u}r Angewandte Mathematik und Mechanik},
	volume={94},
	number={5},
	pages={421-447},
	year={2014},
	publisher={Wiley Online Library}
}

@book{dummit2004abstract,
	title={Abstract algebra},
	author={Dummit, David Steven and Foote, Richard M and others},
	volume={3},
	year={2004},
	publisher={Wiley Hoboken}
}

@book{gallian2021contemporary,
	title={Contemporary abstract algebra},
	author={Gallian, Joseph},
	year={2021},
	publisher={Chapman and Hall/CRC}
}

@article{10.1063/1.1708514,
	author = {Brinkman, W. and Elliott, R. J.},
	title = {Space Group Theory for Spin Waves},
	journal = {Journal of Applied Physics},
	volume = {37},
	number = {3},
	pages = {1457-1459},
	year = {1966},
	month = {03},
	issn = {0021-8979},
}

@article{doi:10.1098/rspa.1966.0211,
	author = {Brinkman, W. F.  and Elliott, Roger James },
	title = {Theory of spin-space groups},
	journal = {Proceedings of the Royal Society of London. Series A. Mathematical and Physical Sciences},
	volume = {294},
	number = {1438},
	pages = {343-358},
	year = {1966},
}

@book{Hall2015,
	title = {Lie Groups,  Lie Algebras,  and Representations: An Elementary Introduction},
	ISBN = {9783319134673},
	ISSN = {2197-5612},
	url = {http://dx.doi.org/10.1007/978-3-319-13467-3},
	DOI = {10.1007/978-3-319-13467-3},
	journal = {Graduate Texts in Mathematics},
	publisher = {Springer International Publishing},
	author = {Hall,  Brian C.},
	year = {2015}
}

@book{Brcker1985,
	title = {Representations of Compact Lie Groups},
	ISBN = {9783662129180},
	ISSN = {0072-5285},
	url = {http://dx.doi.org/10.1007/978-3-662-12918-0},
	DOI = {10.1007/978-3-662-12918-0},
	journal = {Graduate Texts in Mathematics},
	publisher = {Springer Berlin Heidelberg},
	author = {Br\"{o}cker,  Theodor and tom Dieck,  Tammo},
	year = {1985}
}

@book{Tung1985,
	title = {Group Theory in Physics: An Introduction to Symmetry Principles,  Group Representations,  and Special Functions in Classical and Quantum Physics},
	ISBN = {9789812384980},
	url = {http://dx.doi.org/10.1142/0097},
	DOI = {10.1142/0097},
	publisher = {WORLD SCIENTIFIC},
	author = {Tung,  Wu-Ki},
	year = {1985},
	month = aug 
}

@article{DZYALOSHINSKY1958241,
	title = {A thermodynamic theory of "weak" ferromagnetism of antiferromagnetics},
	journal = {Journal of Physics and Chemistry of Solids},
	volume = {4},
	number = {4},
	pages = {241-255},
	year = {1958},
	issn = {0022-3697},
	author = {I. Dzyaloshinsky},
}

@article{PhysRev.120.91,
	title = {Anisotropic Superexchange Interaction and Weak Ferromagnetism},
	author = {Moriya, T\^oru},
	journal = {Phys. Rev.},
	volume = {120},
	issue = {1},
	pages = {91-98},
	numpages = {0},
	year = {1960},
	month = {Oct},
	publisher = {American Physical Society},
}

@article{olive2013symmetry,
	title={Symmetry classes for even-order tensors},
	author={Olive, Marc and Auffray, Nicolas},
	journal={Mathematics and Mechanics of Complex Systems},
	volume={1},
	number={2},
	pages={177-210},
	year={2013},
	publisher={Mathematical Sciences Publishers}
}
	
\end{document}